\documentstyle[prl,aps]{revtex}
\input BoxedEPS.tex
\SetOzTeXEPSFSpecial
\HideDisplacementBoxes

\begin{document}

%\draft{}
\twocolumn[\hsize\textwidth\columnwidth\hsize
           \csname @twocolumnfalse\endcsname
\title{Bulk experimental evidence of half-metallic ferromagnetism in
doped manganites}

\author{Guo-meng Zhao and H. Keller}
\vspace{1cm}
\address{Physik-Institut der Universit\"at Z\"urich, CH-8057 Z\"urich,
Switzerland}
\maketitle
%\date{\today}
\maketitle
%\vskip0.5truecm
\noindent
\begin{abstract}
We report precise
measurements and quantitative data analysis on the low-temperature
resistivity of several ferromagnetic manganite films. We clearly
show that there exists a $T^{4.5}$ term in low-temperature
resistivity, and that this term is in quantitative
agreement with the quantum theory of two-magnon scattering for half
metallic ferromagnets. Our present
results provide the first bulk experimental evidence of half-metallic
ferromagnetism in doped manganites.
\end{abstract}
%PACS numbers:
%\newpage
\vspace{0.4cm}

]
\newpage
%\narrowtext

The concept of half-metallic ferromagnets was first introduced by de
Groot {\em et al.} in 1983 \cite{Groot}. Half-metallic ferromagnets are
characterized by completely
spin-polarized electronic density of states at the Fermi level, that
is, the majority spin channel is metallic while the Fermi energy falls
in a band gap in the minority spin density of states. Such a novel physical
property makes the materials very promising in technological
applications such as single-spin electron sources and high-efficiency magnetic
sensors. The half-metallic feature has been
predicted for CrO$_{2}$
\cite{Schwarz}. However,
spin-resolved photoemission measurement on the material has not
confirmed this prediction \cite{Kamper}. On the other hand, the local density
approximation (LDA) band-structure calculation on
La$_{2/3}$Ca$_{1/3}$MnO$_{3}$ \cite{Pickett} has shown an electronic
structure of a nearly half-metallic ferromagnet (i.e., the majority
spin channel is metallic while the Fermi energy lies within a band
edge of the minority spin channel). Spin-resolved photoemission study
on a La$_{0.7}$Sr$_{0.3}$MnO$_{3}$ film has demonstrated an
electronic structure with spin polarization of nearly 100$\%$ at Fermi
level, which
possibly manifests the half-metallic feature \cite{Park}. However, the
spin-resolved tunneling
measurements showed 54$\%$ (Ref.~\cite{Lu}) and 81$\%$ (Ref.~\cite{Sun})
polarization of conduction electrons in La$_{0.67}$Sr$_{0.33}$MnO$_{3}$. Since
both photoemission and tunneling experiments are rather surface
sensitive, the controversial conclusions drawn from these experiments
are not so surprising. One needs to look for bulk sensitive
experiments to unambiguously demonstrate whether the doped manganites
are truly half-metallic ferromagnets or not. The clarification of this
issue can place an essential constraint on the future prospect of this
material in technological applications.

Since the Fermi level only crosses the majority spin bands in
half-metallic ferromagnets, the
emission or absorption of single magnon is forbidden, because the
carriers have no conducting minority states at low energy to spin-flip
scatter into. Therefore, the $T^{2}$ temperature-dependent term in
the low-temperature resistivity due to single-magnon scatter
\cite{Mannari} should
be absent. Instead, two-magnon scattering is allowed, and leads to
a $T^{4.5}$ temperature dependence in resistivity,
as predicted by Kubo and Ohata \cite{Kubo}. Since this $T^{4.5}$ term
should be rather small, it is difficult to
identify this term if the other contributions to the resistivity are
dominant. Nevertheless, if one can unambiguously identify the $T^{4.5}$
term, one provides bulk experimental evidence for the half-metallic
nature. In this letter we not only show that there indeed exists the $T^{4.5}$
term in several manganite ferromagnets, but also demonstrate that the
coefficient of the $T^{4.5}$
term is in quantitative agreement with the quantum theory of two-magnon
scattering \cite{Kubo}.

In half-metallic ferromagnets, the low-temperature resistivity due to
two-magnon scattering was found to be $\rho_{KO} =
AT^{4.5}$ (Ref.~\cite{Kubo}). The coefficient $A$ has an
analytical expression
in the case of a simple parabolic conduction band (occupied by
single-spin holes) \cite{Kubo}. In terms of the hole density per cell $n$,
the average spin stiffness $\bar{D}$, and the effective hopping integral
$t^{*}$,  the coefficient $A$ can be written as \cite{Kubo}
\begin{eqnarray}
&A =& (\frac{3a\hbar}{32\pi e^{2}})(2 - n/2)^{-2}(6\pi^{2}n)^{5/3}(2.52 +
0.0017\frac{\bar{D}}{a^{2}t^{*}})
\nonumber \\
& &\{\frac{a^{2}k_{B}}{\bar{D}(6\pi^{2})^{2/3}(0.5^{2/3}-n^{2/3})}\}^{9/2}.
\end{eqnarray}
Here we have used the relations: $ak_{F}= (6\pi^{2}n)^{1/3}$ (where
$\hbar k_{F}$ is the Fermi momentum, and $a$ is the lattice constant);
$E_{F}= t^{*}(6\pi^{2})^{2/3}(0.5^{2/3}-
n^{2/3})$ (where the Fermi energy $E_{F}$ is measured from the band
center); the effective spin $S^{*} = 2 - n/2$. The value of $t^{*}$ can be
estimated to be about 40 meV
from the measured effective plasma frequency $\hbar\Omega_{p}^{*}$ =
1.1 eV and $n \sim$ 0.3 in La$_{0.7}$Ca$_{0.3}$MnO$_{3}$ \cite{Simpson}.
In ferromagnetic manganites,  $\bar{D}$ is
about 100 meV \AA$^{2}$ (see below), so the term
0.0017$\bar{D}/a^{2}t^{*}$$<$$<$ 2.52, and can be dropped out in
Eq.~1. Then there are two parameters $n$ and
$\bar{D}$ that determine the magnitude of $A$. In doped manganites, $n$
should be approximately equal to the doping level $x$, as it is the
case in La$_{1-x}$Sr$_{x}$MnO$_{3}$ system (see below).
The average spin stiffness $\bar{D}$ should be close to the long-wave
spin stiffness $D(0)$ if there is
negligible magnon softening near the zone boundary.

It has recently been shown \cite{ZhaoPRL00}  that
the dominant contribution to the
low-temperature resistivity is due to scattering from a soft
optical phonon mode, which gives a term proportional to
$\omega_{s}/\sinh^{2}(\hbar\omega_{s}/2k_{B}T)$, where $\omega_{s}$ is
the frequency of a soft optical
mode. If we include a possible contribution from two-magnon
scattering \cite{Kubo}, or from acoustic-phonon scattering
\cite{Mannari}, the temperature dependent part of the
resistivity can be generally expressed as
\begin{equation}
\rho(T) - \rho_{o} = AT^{\alpha}+
B\omega_{s}/\sinh^{2}(\hbar\omega_{s}/2k_{B}T),
\end{equation}
where $\rho_{o}$ is the residual resistivity; $A$ and $B$ are
temperature independent coefficients. The power $\alpha$ = 4.5 if the
major contribution is from two-magnon scattering,
while $\alpha$ = 5 if the main contribution is due to acoustic phonon
scattering. We can make a distinction between the two cases. For
acoustic phonon scattering, the coefficient $A$ should be independent
of the applied magnetic field if $m^{*}/n$ or $\rho_{o}$ does not depend
on the field (where $m^{*}$ is the effective mass of
carriers). This is because $A$ $\propto$
$(\frac{m^{*}}{n})^{2}(\theta_{D})^{-5}$ in this case, where $\theta_{D}$
is the
Debye temperature \cite{Mannari}. For two-magnon scattering,
a magnetic field induces a
gap in spin-wave excitations, so that $A$ should decrease with
increasing magnetic field.
\begin{figure}[htb]
    \ForceWidth{7cm}
	\centerline{\BoxedEPSF{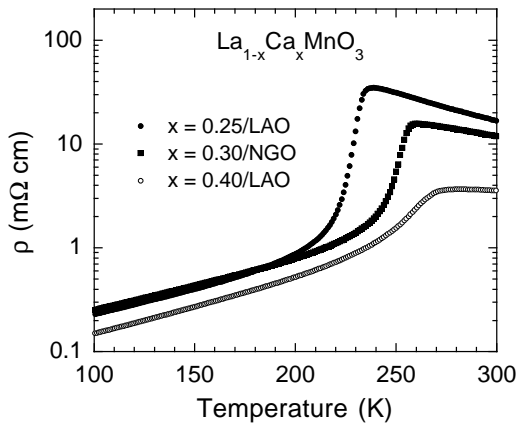}}
	\vspace{0.3cm}
	\caption[~]{The resistivity of the films
La$_{1-x}$Ca$_{x}$MnO$_{3}$ over
	100-300 K. As $x$ increases, the metal-insulator crossover
	temperature increases. }
	\protect\label{Fig.1}
\end{figure}

Epitaxial thin films of La$_{1-x}$Ca$_{x}$MnO$_{3}$ were grown
on $<$100$>$ LaAlO$_{3}$ (LAO) or NdGaO$_{3}$ (NGO) single crystal
substrates by pulsed laser
deposition using a KrF excimer laser \cite{Prellier}.
The films were post annealed in 1 bar oxygen at 940 $^{\circ}$C for
10 h.
The resistivity was measured using the van der Pauw technique, and the
contacts were made by silver paste. The measurements were
carried out in a Quantum Design measuring system.

Fig.~1 shows the temperature dependence of the resistivity for the films of
La$_{1-x}$Ca$_{x}$MnO$_{3}$
over 100-300 K. As $x$ increases the metal-insulator crossover
temperature increases. The Curie temperature $T_{C}$
should coincide with a
temperature where $d\ln\rho/dT$ exhibits a maximum.
The $T_{C}$ values with this definition are listed in Table I.

In Fig.~2, we show the temperature-dependent part of
the resistivity at low temperatures
for La$_{0.75}$Ca$_{0.25}$MnO$_{3}$ film measured in zero magnetic
field (a), and in a magnetic field of 4 T (b).  We fit the data in
zero field  by
Eq.~2 using
three fitting parameters: $A$, $B$ and $\omega_{s}$, and a fixed
$\alpha$ = 5. Meanwhile, we fit the data in 4 T magnetic field with
two fitting parameters:
$A$ and $B$, and with fixed $\omega_{s}$ and $\alpha$ which are the same
as those in zero field case. One can see that the fits are very good
in both cases. Note that we have excluded
the data above 100 K in
the fitting since $n/m^{*}$ above 100 K becomes temperature
dependent \cite{Simpson}.
\begin{figure}[htb]
    \ForceWidth{7cm}
	\centerline{\BoxedEPSF{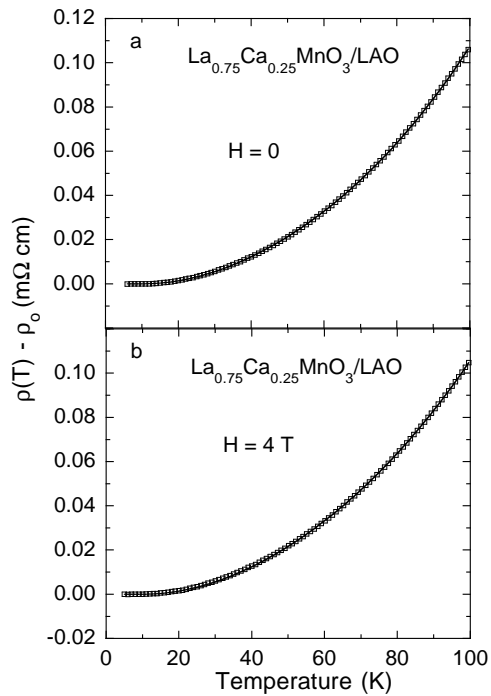}}
	\vspace{0.3cm}
	\caption[~]{The temperature-dependent part of the resistivity at low
	temperatures for the film of
La$_{0.75}$Ca$_{0.25}$MnO$_{3}$ measured in zero magnetic field (a),
and in 4 T magnetic field (b). The solid lines are
the fitted curves by Eq.~2 with a fixed $\alpha$ = 5. The residual
resistivity $\rho_{o}$ does not depend on the magnetic field within
the experimental uncertainty. }
	\protect\label{Fig.2}
\end{figure}

An important finding from the fits is that the parameters $B$ and
$\omega_{s}$ are independent of magnetic field, while
the value of the parameter $A$ in 4 T magnetic field is smaller than that
in zero
magnetic field by a factor of 1.5. Such a strong field dependence of
the coefficient $A$ is not expected from acoustic phonon scattering
unless $m^{*}/n$ or $\rho_{o}$ also strongly depends
on the field \cite{Mannari}. The fact that $\rho_{o}$ is independent
of the magnetic field rules out the possibility that the $AT^{\alpha}$ term
in Eq.~2 can arise from acoustic phonon scattering.

Fig.~3 shows the temperature-dependent part of
the resistivity at low temperatures
in zero magnetic field for La$_{1-x}$Ca$_{x}$MnO$_{3}$ films with
different $x$. The solid
lines are the fitted curves by Eq.~2 with three fitting parameters: $A$,
$B$ and
$\omega_{s}$, and with a fixed $\alpha$ = 4.5. The values of the
fitting parameter $A$ are
summarized in Table I.  The $\hbar\omega_{s}$ value is
6.41$\pm$0.03 meV for $x$ = 0.25, 7.40$\pm$0.06 meV for $x$ =
0.30, and 6.90$\pm$0.06 meV for $x$ =
0.40. The doping dependence of $\omega_{s}$ is very simialr to that
of $\theta_{D}$ found for La$_{1-x}$Sr$_{x}$MnO$_{3}$ system \cite{Okuda}.
The absolute values of $\hbar\omega_{s}$ are very close to the phonon
energy (6 meV) of the rotational
vibrations of the oxygen octahedra in a similar
perovskite Ba(Pb$_{0.75}$Bi$_{0.25}$)O$_{3}$ \cite{Reichardt}. This
rotational mode is shown to be strongly coupled to conduction
electrons by tunneling experiment \cite{Reichardt}.
\begin{figure}[htb]
    \ForceWidth{7cm}
	\centerline{\BoxedEPSF{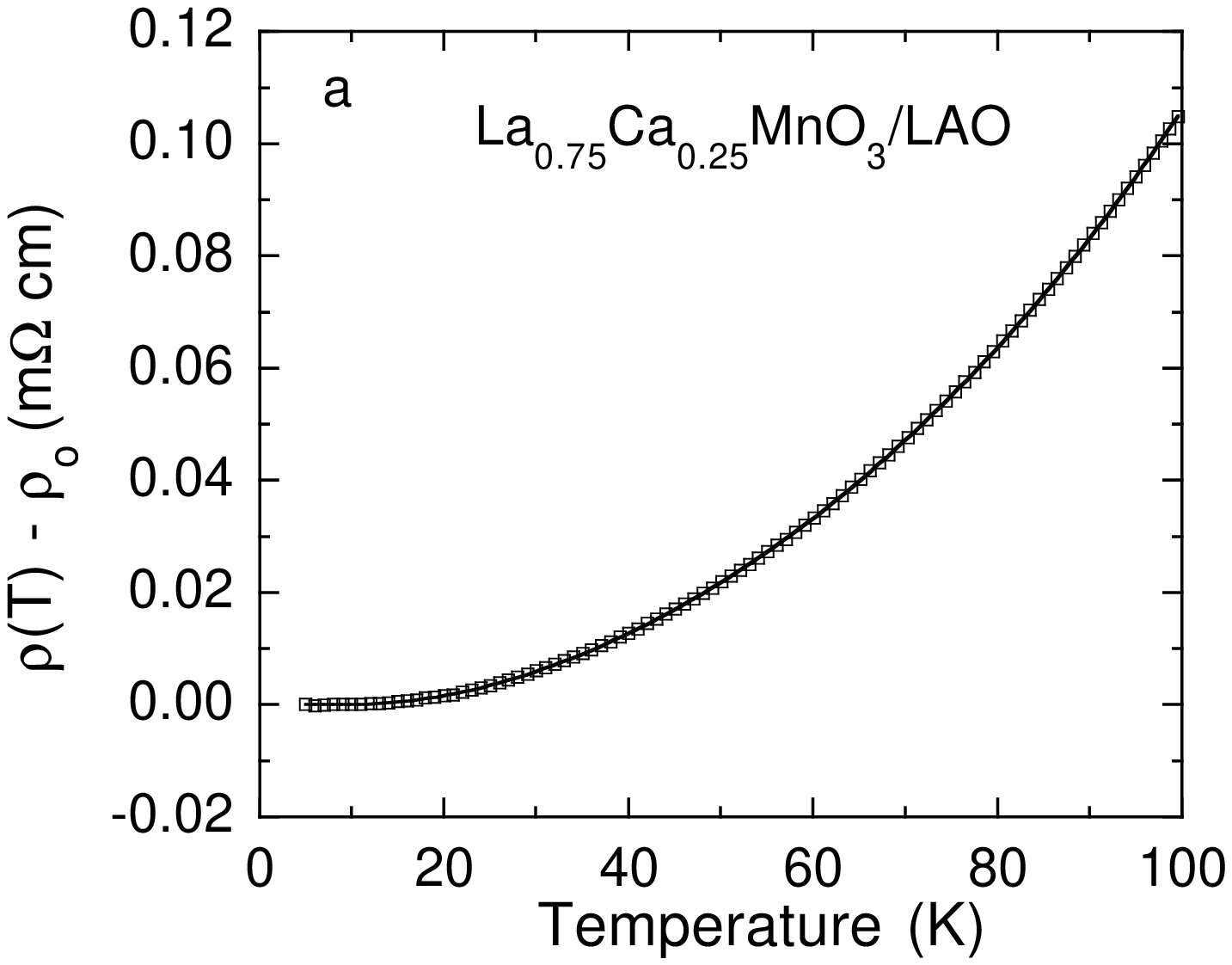}}
	\ForceWidth{7cm}
	\centerline{\BoxedEPSF{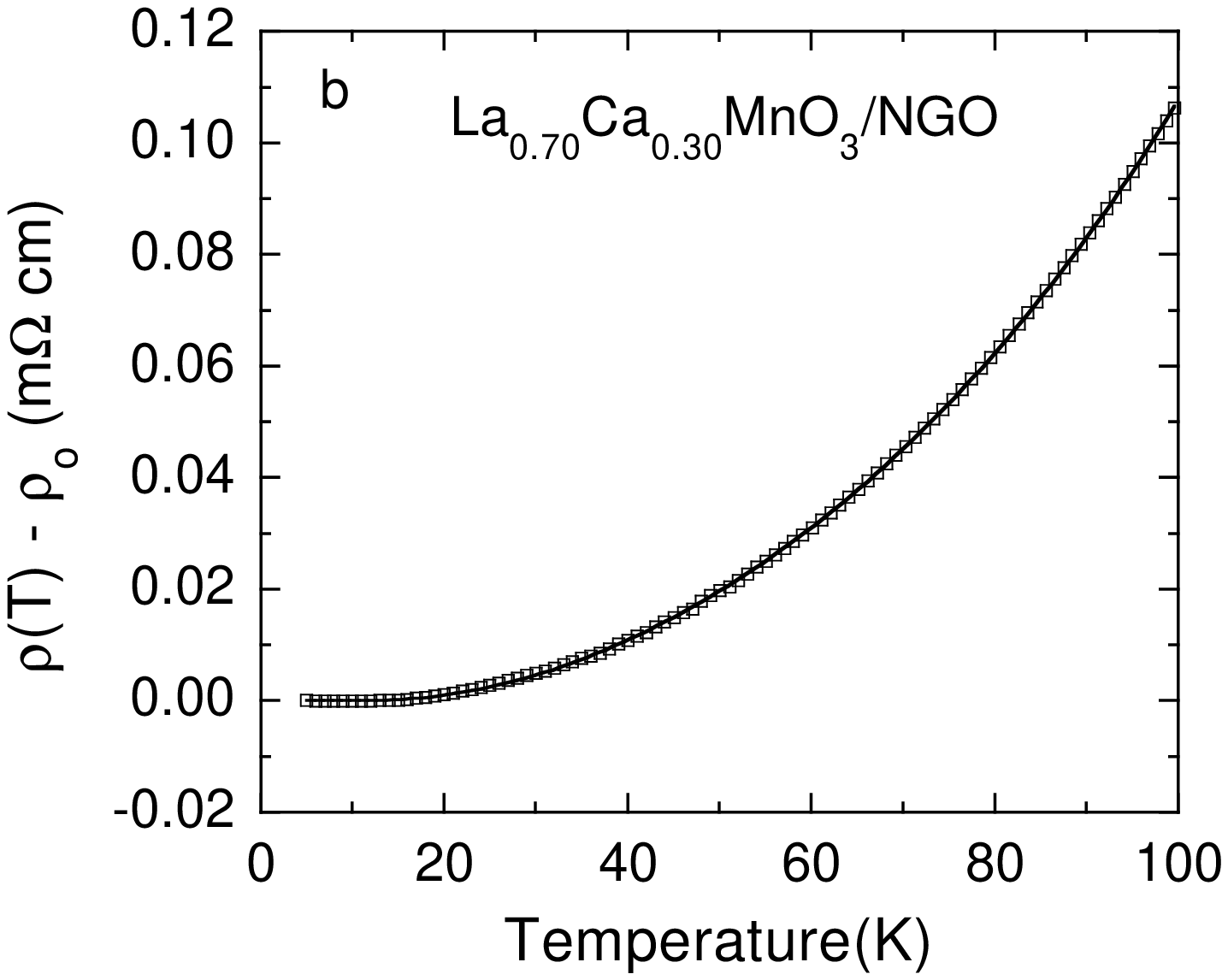}}
	\ForceWidth{7cm}
	\centerline{\BoxedEPSF{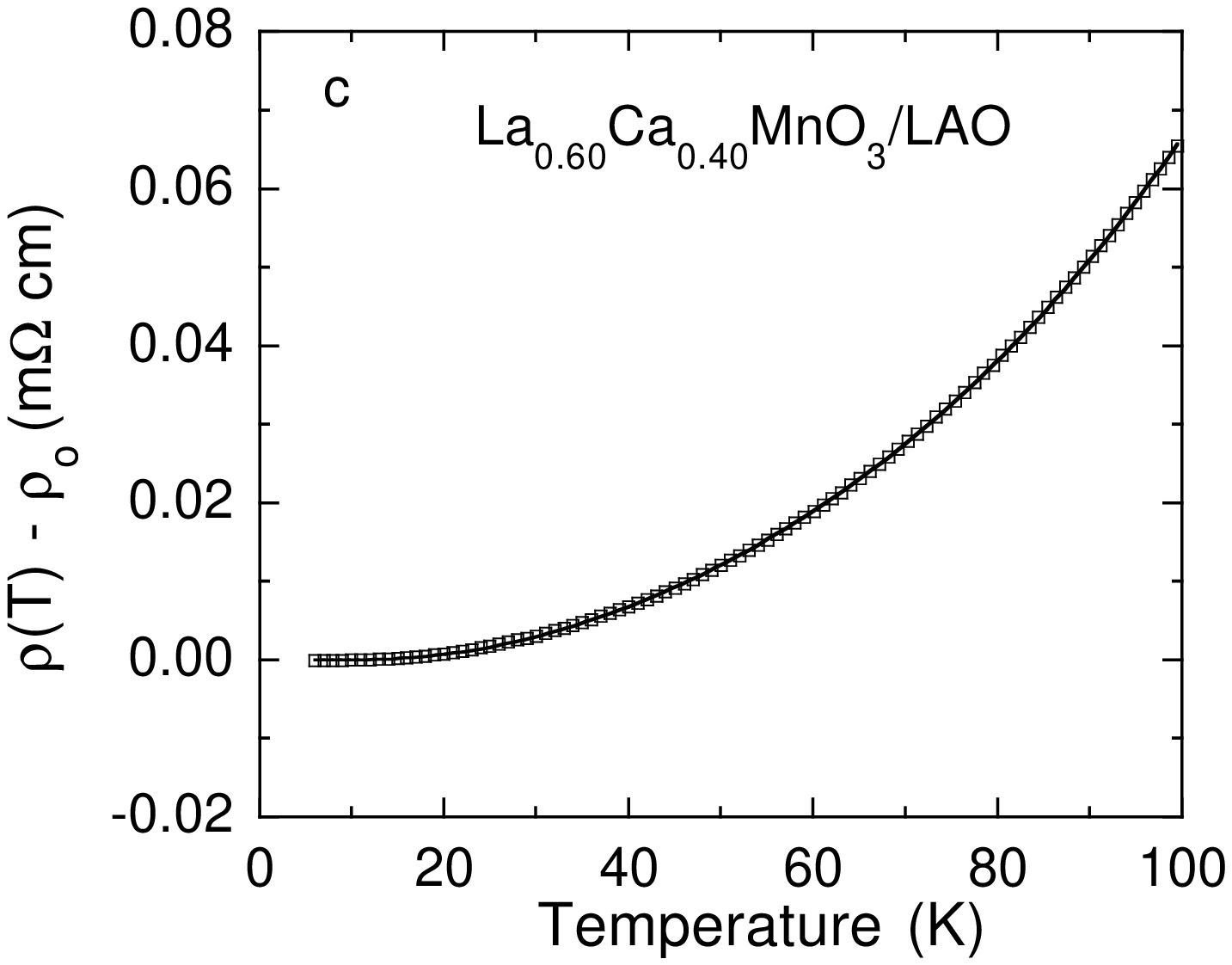}}
	\vspace{0.6cm}
	\caption[~]{The temperature-dependent part of the resistivity at low
	temperatures for the
La$_{1-x}$Ca$_{x}$MnO$_{3}$ films with different $x$.  The solid lines are
fitted curves by Eq.~2 with a fixed $\alpha$ = 4.5.}
	\protect\label{Fig.3}
\end{figure}
\begin{figure}[htb]
    \ForceWidth{7cm}
	\centerline{\BoxedEPSF{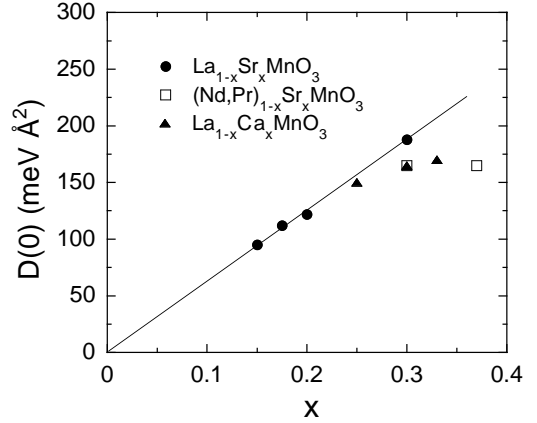}}
	\vspace{0.3cm}
	\caption[~]{The long-wave spin stiffness $D(0)$ as a
function of $x$ for different manganite systems. The data of
La$_{1-x}$Sr$_{x}$MnO$_{3}$ are taken from Ref.~\cite{Hirota,Doloc}. The data
of La$_{1-x}$Ca$_{x}$MnO$_{3}$ are from
Ref.~\cite{Dai,Lynn}. The data of (Nd,Pr)$_{1-x}$Sr$_{x}$MnO$_{3}$
are taken from Ref.~\cite{Baca}.}
	\protect\label{Fig.4}
\end{figure}

In order to further confirm that the $T^{4.5}$ term indeed originates from
two-magnon scattering, it is neccessary to show that
the values of the coefficient $A$ in Table I
should be in quantitative
agreement with  Eq.~1. Since the $A$ value in Eq.~1 is sensitive to
the hole density $n$, it is essential to determine the $n$ value more
reliably. Roughly speaking, one might expect that $n$ $\simeq$ $x$. Since
doped holes mainly reside on the oxygen sites \cite{Saitoh,Ju}, each
oxygen hole will
produce a ferromagnetically coupled bond with an
exchange energy $J$$\sim$$t_{pd}^{4}/\Delta^{3}$, where $t_{pd}$
is a hybridization
matrix element between the $d$ and $p$ orbitals, and $\Delta$ is a charge
transfer gap \cite{Khomskii}. Then one can readily show that the long-wave
spin stiffness $D(0) \propto n$. In Fig.~4, we plot $D(0)$ as a
function of $x$ for different manganite systems. For
La$_{1-x}$Sr$_{x}$MnO$_{3}$ system, the linear relation between
$D(0)$ and $x$ holds up to $x$ = 0.3, implying that $n$ = $x$ for
0.15 $\leq$ $x$ $\leq$ 0.3. For La$_{1-x}$Ca$_{x}$MnO$_{3}$ and
(Nd,Pr)$_{1-x}$Sr$_{x}$MnO$_{3}$ systems, $D(0)$ starts to deviate
\begin{table}[htb]
	\caption[~]{The summary of the fitting parameter $A$, the measured
	$T_{C}$, the estimated hole density per cell $n$,
	and the calculated $\bar{D}/k_{B}T_{C}$ for
	La$_{1-x}$Ca$_{x}$MnO$_{3}$. The uncertainty in $T_{C}$ is $\pm$ 1
	K. The measured values of $T_{C}$ and $D(0)/k_{B}T_{C}$ from
	neutron scattering
	for La$_{0.7}$Sr$_{0.3}$MnO$_{3}$ (LSMO30) \cite{Martin} are also
included in the last row. }
	\begin{center}
    \begin{tabular}{lccccc}
    	Compounds &$T_{C}$ & $n$ &$A$ & $\frac{\bar{D}}{k_{B}T_{C}}$&
    	$\frac{D(0)}{k_{B}T_{C}}$
    	 \\
    	 $x$&(K) & &(m$\Omega$cm/K$^{4.5}$)&
         (\AA$^{2}$)& (\AA$^{2}$)\\
    	\hline
    	0.25 (H = 0) &232 &0.24 &1.20(2)$\times$10$^{-11}$ &
    	5.97& \\
    	0.25 (H = 4 T)& & &0.82(2)$\times$10$^{-11}$ &\\
    	0.30&253 &0.26 &1.70(3)$\times$10$^{-11}$ &
    	5.71& \\
    0.40 &262 &0.26&1.27(3)$\times$10$^{-11}$ & 5.88&\\
    LSMO30 &378 & & & & 5.8(2)\\
\end{tabular}
\end{center}
\protect\label{Tab1}
\end{table}
\noindent
from the
linear relation above $x$ = 0.2 and remains a constant for 0.30
$\leq$ $x$ $\leq$ 0.40. From the linear relation between $n$ and $D(0)$,  we
find $n$ = 0.24 for $x$ = 0.25, $n$ = 0.26
for $x$ = 0.30 and 0.40.

Now we can calculate the average spin stiffness $\bar{D}$ using Eq.~1 and
the $A$ values listed in Table I. The calculated $\bar{D}/k_{B}T_{C}$ values
for three manganite compounds are summarized in Table I.
It is remarkable that the $\bar{D}/k_{B}T_{C}$ values deduced from the
resistivity data are very close to the $D(0)/k_{B}T_{C}$ value
(5.8$\pm$0.2 \AA$^{2}$) for
La$_{0.7}$Sr$_{0.3}$MnO$_{3}$ from neutron scattering \cite{Martin}.
Since the magnon
softening near the zone boundary is negligible when the $T_{C}$
is higher than 350 K
\cite{Hwang}, there should a negligible magnon softening in the
compound La$_{0.7}$Sr$_{0.3}$MnO$_{3}$ with the highest $T_{C}$ = 378
K. In this case, one should expect that
$\bar{D}$ $\simeq$ $D(0)$. On the other hand, $\bar{D}$ $<$ $D(0)$ if
there is a magnon softening near the zone boundary as the case of
low $T_{C}$ materials \cite{Hwang}. In any cases, one might expect
that the average $\bar{D}$ should be proportional to $T_{C}$ so that
$\bar{D}/k_{B}T_{C}$ is a univeral constant in the manganite system.
This has indeed been verified by the result shown in Table I. Therefore, our
present results provide a quantitative confirmation for
the quantum theory of two-magnon
scattering.

Since the quantum theory of two-magnon scattering is valid only for
half-metallic ferromagnets, the quantitative proof for the theory in
the doped manganites gives bulk experimental evidence that
the ferromagnetic manganites are indeed half-metallic materials. We
are not aware of any other materials which have been confirmed to be
half-metallic ferromagnets by bulk-sensitive experiments.
The unique half-metallic nature only found in these
doped manganites makes them one of the most important and useful materials
in future technological applications.

In summary, we report precise
measurements and quantitative data analysis on the low-temperature
resistivity of several ferromagnetic manganite films. We
show that there exists a $T^{4.5}$ term in low-temperature
resistivity, and that this term is in quantitative
agreement with the quantum theory of two-magnon scattering for half
metallic ferromagnets. Our present
results provide the first bulk experimental evidence of half-metallic
ferromagnetism in doped manganites.

{\bf Acknowledgement}: We would like to thank W. Prellier and D. J.
Kang for providing high-quality manganite films. The work was supported by
the Swiss National Science
Foundation.

\bibliographystyle{prsty}

\begin{thebibliography}{10}
\bibitem{Groot} R. A. de Groot, F. M. M\"uller, P. G. van Engen, and
K. H. J. Buschow, Phys. Rev. Lett.
\textbf{50}, 2024 (1983).
\bibitem{Schwarz}K. Schwarz, J. Phys. F \textbf{16}, L211 (1986).
\bibitem{Kamper}K. P. K\"amper, W. Schmitt, G. G\"untherodt, R. J. Gambin,
and R. Ruf, Phys. Rev. Lett. \textbf{59}, 2788 (1988).
\bibitem{Pickett}W. E. Pickett and D. J. Singh, Phys.
Rev. B \textbf{53}, 1146 (1996).
\bibitem{Park}J. H. Park, E. Vescovo, H.-J. Kim, C. Kwon, R. Ramesh,
and T. Venkatesan, Nature (London) \textbf{392}, 794 (1998).
\bibitem{Lu}Y. Lu {\em et al.},  Phys.
Rev. B \textbf{54}, R8357 (1996).
\bibitem{Sun}J. Z. Sun, L. Krusin-Elbaum, P. R. Duncombe, A. Gupta,
and R. B. Laibowwitz, Appl. Phys. Lett. \textbf{70}, 1769 (1997).
\bibitem{Mannari}I. Mannari, Prog. Theor.  Phys. \textbf{22}, 335 (1959).
\bibitem{Kubo}K. Kubo and N. A.  Ohata, J. Phys. Soc. Jpn. \textbf{33},
21 (1972).
\bibitem{Simpson}J. R. Simpson, H. D. Drew, V. N. Smolyninova, R. L.
Greene, M. C. Robson, A. Biswas, and M.  Rajeswari, Phys. Rev. B \textbf{60},
R16 263 (1999).
\bibitem{ZhaoPRL00}G. M. Zhao, V. Smolyaninova, W. Prellier, and H.
Keller, Phys. Rev. Lett.
\textbf{84}, 6086 (2000).
\bibitem{Prellier}W. Prellier, M. Rajeswari, T. Venkatesan, and R.
L. Greene,  Appl. Phys. Lett. \textbf{75},
1446 (1999).
\bibitem{Okuda}T. Okuda, A. Asamitsu, Y. Tomioka, T. Kimura, Y.
Taguchi, and Y. Tokura,  Phys. Rev.
Lett. \textbf{81}, 3202 (1998).
\bibitem{Reichardt}W. Reichardt, B. Batlogg, and J. P. Remeika,
Physica B \textbf{135}, 501 (1985).
\bibitem{Saitoh}T. Saitoh, A. E. Bocquet, T. Mizokawa, H.
Namatame, A.  Fujimori, M. Abbate, Y. Takeda, and M. Takano,  Phys.
Rev. B \textbf{51}, 13942 (1995).
\bibitem{Ju}H. L. Ju, H. C. Sohn, and K. M. Krishnan,  Phys. Rev. Lett.
\textbf{79}, 3230 (1997).
\bibitem{Khomskii}D. I. Khomskii and G. A. Sawatzky,
Solid State Commun. \textbf{102}, 87 (1997).
\bibitem{Hirota}K. Hirota, N. Kaneko, Y. Endoh, M. C. Martin, and G.
Shirane,  Physica B
\textbf{237-238}, 36 (1997).
\bibitem{Doloc}L. Vasiliu-Doloc, J. W. Lynn, A. H. Moudden, A. M. de
Leon-Guevara, and A. Revcolevschi,  Phys.
Rev. B \textbf{58}, 14913 (1998).
\bibitem{Dai}P. Dai, H. Y. Hwang, J. Zhang, J. A. Fernandez-Baca, S.-W.
Cheong,
C. Kloc, Y. Tomioka, and Y. Tokura, Phys.
Rev. B \textbf{61}, 9553 (2000); $D(0)$ = 150 meV \AA$^{2}$ for
La$_{0.75}$Ca$_{0.25}$MnO$_{3}$ (unpublished data).
\bibitem{Lynn}J. W. Lynn, R. W. Erwin, J. A. Borchers, Q. Huang, A.
Santoro, J. L. Peng, and Z. Y. Li, Phys. Rev.
Lett. \textbf{76},
4046 (1996).
\bibitem{Baca}J. A. Fernandez-Baca, P. Dai, H. Y. Hwang, C. Kloc, and
S.-W. Cheong, Phys. Rev.
Lett. \textbf{80},
4012 (1998).
\bibitem{Martin}M. C. Martin, G. Shirane, Y. Endoh, K. Hirota, Y.
Moritomo, and Y. Tokura, Phys.
Rev. B \textbf{53}, 14 285 (1996).
\bibitem{Hwang}H. Y. Hwang, P. Dai, S.-W. Cheong, G. Aeppli, D. A.
Tennant, and H. A. Mook, Phys. Rev. Lett.
\textbf{80}, 1316 (1998).







\end{thebibliography}

\end{document}